%% file: MMstab2.tex
\documentclass[a4paper,11pt]{article}
\pdfoutput=1 % if your are submitting a pdflatex (i.e. if you have
\usepackage{jcappub} 

\usepackage[T1]{fontenc} % if needed

\usepackage{graphicx}

\usepackage{rotating}

\usepackage{dcolumn}

\usepackage{amssymb}

\usepackage{amsmath}

\usepackage{bm}% bold math

\usepackage{listings}

\usepackage{fancyhdr}

\usepackage{wrapfig}

\usepackage{subfig}

\usepackage{color}

\include{mydefs}

\title{\boldmath Stability analysis and future singularity of the $m^2 R \bo^{-2} R$ model of non-local gravity}

\author{Yves Dirian,}
\author{Ermis Mitsou}
\affiliation{D\'epartement de Physique Th\'eorique and Center for Astroparticle Physics,\\ Universit\'e de Gen\`eve, 24 quai Ansermet, CH--1211 Gen\`eve 4, Switzerland}

\emailAdd{yves.dirian@unige.ch}
\emailAdd{ermis.mitsou@unige.ch}

\abstract{We analyse the classical stability of the model proposed by Maggiore and Mancarella, where gravity is modified by a term $\sim m^2 R \bo^{-2} R$ to produce the late-time acceleration of the expansion of the universe. Our study takes into account all excitations of the metric that can potentially drive an instability. There are some subtleties in identifying these modes, as a non-local field theory contains dynamical fields which yet do not correspond to degrees of freedom. Since some of them are ghost-like, we clarify the impact of such modes on the stability of the solutions of interest that are the flat space-time and cosmological solutions. We then find that flat space-time is unstable under scalar perturbations, but the instability manifests itself only at cosmological scales, i.e. out of the region of validity of this solution. It is therefore the stability of the FLRW solution which is relevant there, in which case the scalar perturbations are known to be well-behaved by numerical studies. By finding the analytic solution for the late-time behaviour of the scale factor, which leads to a big rip singularity, we argue that the linear perturbations are bounded in the future because of the domination of Hubble friction. In particular, this effect damps the scalar ghost perturbations which were responsible for destabilizing Minkowski space-time. Thus, the model remains phenomenologically viable.}

\begin{document}
\maketitle
\flushbottom

\section{Introduction and summary}

Recently, Maggiore and Mancarella (MM) proposed a model based on the following non-local modification of General Relativity (GR) \cite{MM}, \footnote{Here $d$ is the space dimension, we work with a signature of mostly pluses and the convention $R_{\mu\nu} = \pa_{\ro} \Ga^{\ro}_{\,\,\,\mu\nu} - \dots$.} 
\beq \label{eq:MM}
S_{\rm MM} = \frac{1}{16 \pi G} \int \ed^{d+1} x\, \sqrt{-g} \[ R - \frac{1}{2}\, \al^2  R\, \frac{m^2}{\bo^2} R \]  \, , \hspace{1cm} \al \equiv \sqrt{\frac{d-1}{2d}} \, ,
\eeq
in order to reproduce the observed amount of dark energy. Here the $\bo^{-1}$ operator is a formal inverse of $\bo$ in the scalar representation which can be expressed as the convolution with a bi-scalar Green's function $G$, \footnote{Incidentally, since we consider no homogeneous solution in the definition of $\bo^{-1}$, we have that $\bo^{-1} 0 = 0$. \label{fn:ztoz}}
\beq \label{eq:bodef}
(\bo^{-1} R) (x) \equiv \int \ed^{d+1} y\, \sqrt{-g(y)} \, G(x,y) R(y) \, , \hspace{1cm}  \bo_x G(x, y) = \frac{1}{\sqrt{-g(x)}} \, \de^{(d+1)}(x-y) \, .
\eeq
This model is currently receiving particular attention \cite{DFKKM, CKMT, BLHBP, DFKMP} because its phenomenology seems to privilege it among other non-local models that have been confronted with observations \cite{DW1, DW2, W, WD, K1, K2, DP1, DP2, JMM, MT, M, FMM2, KM, NT}. Here the non-local term is controlled by a mass parameter $m$, in contrast with the non-local models of Deser and Woodard \cite{DW1} and of Barvinsky \cite{B1}, where no new scale is introduced in the theory. A subtlety of non-local (in time) actions is that the corresponding equations of motion are not causal because both retarded and advanced Green's functions appear. The MM model uses the prescription of turning all of them into retarded ones 
\beq \label{eq:retG}
G(x,y) = 0 \, , \hspace{1cm} \mbox{unless $y$ is in the past light-cone of $x$} \, ,
\eeq 
after the variation, thus losing the direct relation with the action (\ref{eq:MM}). The latter then just serves as a compact way of displaying the model's information since the corresponding equations of motion 
\beq \label{eq:EOM}
G_{\mu\nu} + \De G_{\mu\nu}\[ \bo^{-1} \] = 8\pi G\, T_{\mu\nu} \, ,
\eeq
are considerably lengthier. Most importantly, the advantage of deriving these equations from a generally covariant action is that one obtains an automatically conserved correction, i.e. $\na^{\mu} \De G_{\mu\nu} = 0$, a property which is not spoiled by the causality prescription. 

Condition (\ref{eq:retG}) does not entirely determine the Green's function, as one must still impose initial conditions to solve (\ref{eq:bodef}). We therefore invoke some local coordinates $x^{\mu}$, where $\pa_0$ is time-like and $\pa_i$ are space-like. The MM model is understood as an effective theory valid below some energy scale and thus only after some time $t_0$ in the cosmological history. Demanding that the non-local effect starts at $t_0$ means that we must impose the boundary conditions\footnote{Note that this choice is coordinate-dependent, or alternatively, that the choice of such a $G$ privileges a coordinate system: it is the system in which the initial conditions take the simplest form (\ref{eq:Gfchoice}).}
\beq \label{eq:Gfchoice}
\left. G(x,y) \right|_{x^0 = t_0} = 0  \, , \hspace{1cm} \left. \pa_0 G(x,y) \right|_{x^0 = t_0} = 0 \, .
\eeq  
Equations (\ref{eq:retG}) and (\ref{eq:Gfchoice}) must be appended to (\ref{eq:EOM}) to fully determine the MM model. 

The phenomenology of (\ref{eq:retG})(\ref{eq:EOM})(\ref{eq:Gfchoice}) has shown to be very appealing, especially when taking into account the simplicity of the model (a one-parameter extension of GR along with a Green's function prescription). An important property is the absence of vDVZ discontinuity \cite{MM, KM} which implies that, for small enough $m$, the modification does not affect solar system tests. Since $m$ has to be of the order of the Hubble parameter today $H_0$, small scale gravitation is unaffected indeed. Another nice feature, shared by all $\bo^{-1}R$-based models, is that by choosing $t_0$ to lie well inside the radiation-dominated era (RD) one obtains a natural onset for the appearance of dark energy. Indeed, since $R = 0$ during RD, the deviation from GR starts in the transition from radiation to matter domination (MD) and the precise value of $t_0$ is irrelevant. 

Furthermore, the model has predictive power. Fixing $m$ so that we recover the observed value of the dark energy density, automatically determines the corresponding equation of state parameter for the effective dark energy $w_{\rm DE}$. The latter is found to lie on the phantom side today \cite{MM,DFKKM} and is consistent with the Planck \cite{P} and type Ia Supernovae data \cite{SNe}. Moreover, the linear perturbations in the scalar sector were found to be numerically stable and qualitatively close to the ones in $\La$CDM \cite{DFKKM}. The model fulfils the present constraints required for structure formation, at the linear \cite{DFKKM, BLHBP} and non-linear level \cite{BLHBP}, while still being different enough to be discriminated against $\La$CDM by future observations.

Despite these phenomenological successes, there are however still some structural and conceptual open issues that remain unaddressed. One of these is the question of the classical stability of solutions of physical relevance, such as the Minkowski, Schwarzschild and Friedmann-Lema\^itre-Robertson-Walker (FLRW) space-times. Indeed, the only available information on the subject lies in \cite{KM} and \cite{DFKKM}. In the former the authors showed that, for a Schwarzschild black hole of mass $M_{\rm BH}$, the corrections to the spectrum of its quasi-normal modes is analytic in the ratio $m / M_{\rm BH}$, so that for realistic $M_{\rm BH}$ no instability is expected for the tensor modes\footnote{Actually, they have studied the corrections to the Schwarzschild solution for a different non-local model, but whose relevant equations in that case turn out to be the same as for the MM model \cite{MM}.}. In \cite{DFKKM}, the authors have integrated numerically the linear cosmological perturbations in the scalar sector and showed that they are stable at both super and sub-horizon scales. The aim of the present paper is to perform a more thorough stability analysis which would rely as much as possible on analytical arguments, and which would take into account all possible excitations of $g_{\mu\nu}$ that can potentially drive an instability.  

Following the method of \cite{DW2}, in section \ref{sec:DOF} we perform in a rigorous way the degree of freedom count of the theory to find that it has the same spectrum as GR, that is, only the initial conditions of the gravitational waves are free to choose. In section \ref{sec:GWstab}, following again the methods of \cite{DW2}, we show that the solutions of physical relevance are indeed stable against such excitations. However, this approach, which works directly at the non-local level, does not allow us to say much about whether these solutions may be destabilized by the evolution of other excitations than gravitational waves. 

In section \ref{sec:loc} we introduce the appropriate framework for treating this question which is the ``local'' formulation. As the name suggests, one simply integrates-in auxiliary fields in order to localize (and subsequently diagonalize) the action, so that a clearer picture of the dynamics emerges\footnote{Localization procedures usually involve the use of Lagrange multipliers \cite{MM, DFKKM, JNOSTZ, BNOS, NO1, NO2, CENO, NOSZ, K1, K2, Kosh}. In the case of the MM model, we will see that it is possible to localize in a more economic fashion.}. The resulting action must be treated with care because the auxiliary fields have kinetic terms, but their initial conditions are fixed by the choice of Green's function for $\bo^{-1}$, so they do not represent degrees of freedom of the theory \cite{FMM1, DW2}. Most importantly, it turns out that one of these fields is a ghost, i.e. its kinetic term has the wrong sign\footnote{This was already pointed-out in \cite{MM, DFKKM} for the linearized non-local theory, but here the local formulation allows us to show that it is actually present on {\it any} background, i.e. the ghost cannot condense.}. Despite the fact that this field has ``frozen'' initial conditions, it can still destabilize the solutions of interest because of non-linear effects. For instance, Minkowski space-time is unstable against infrared excitations of that field. Practically however, this is not problematic because the corresponding scales are cosmological, so that Minkowski is not the solution of interest there anyway.

On the other hand, in \cite{DFKKM} the authors have shown numerically that the FLRW solution is stable against {\it all} scalar perturbations at {\it all} scales. Seeing that $H(t)$ grows in the dark energy era (DE), the natural deduction\footnote{See the discussion on page 25 of \cite{DFKKM}.} has been that Hubble friction dominates over any other potentially dangerous effect, thus making all perturbations asymptotically constant. Here we make this statement more rigorous by finding the analytic solution for the scale factor $a(t)$ in the DE era, in section \ref{sec:bigrip}. We find that it exhibits a future singularity, i.e. an infinite $a$ at finite $t$, an example of the so-called ``big rip'' scenario \cite{C, CKW}. This is a generic feature of theories with ghost-like fields, since the corresponding energy-momentum tensor violates the null energy-condition. As a consequence, the Hubble friction {\it must} dominate over any other effect at some point since $H(t)$ blows up at finite time. The perturbations over FLRW therefore do contain a ghost (with frozen initial conditions) but they inevitably end up ``diluted'' by the violent expansion of the universe.  

Finally, let us stress that the MM model is a classical theory since all of its information lies in its equations of motion (\ref{eq:EOM}) and the Green's function prescriptions (\ref{eq:retG}), (\ref{eq:Gfchoice}). After having analysed the above aspects, it is however a natural question to wonder whether there exists a quantum theory having it as its classical limit, or more generally, whether and how this model could be related to a quantum theory. For completeness, we thus give a self-contained and conservative discussion of the quantum issue in the appendix \ref{sec:Qstab}.

\section{Non-local formulation} \label{sec:nonloc}

\subsection{Equations of motion} \label{sec:EOM}

A common feature of non-local actions is the fact that the corresponding equations of motion, derived using the standard variational principle, are necessarily non-causal. Indeed, the variation brings both the non-local operator (here $\bo^{-1}$) {\it and} its transpose\footnote{Which, by the properties of Green's functions, is also an inverse of $\bo$.} in situations such as
\beq
\int \ed^D x\, \ph\, \bo^{-1} \de \psi = \int \ed^D x\, \de \psi \( \bo^{-1} \)^T \ph  \, ,
\eeq
so if the former is retarded (causal) then the latter is advanced. If one is only interested in obtaining causal non-local equations of motion, then the simplest solution is to just turn all Green's functions into retarded ones {\it by hand} at the end of the variation. This is in fact the standard way of proceeding (see \cite{TW} and references therein) and it is the one used for the MM model. Incidentally, both non-local terms $R \bo^{-2} R$, or $(\bo^{-1} R)^2$ in the action will give the same equations of motion and this effectively translates in the freedom of integrating by parts $\bo^{-1}$.

Another subtlety is the fact that $\bo^{-1}$ is by definition a right inverse (\ref{eq:bodef}), i.e. $\bo \bo^{-1} = {\rm id}$, but not a left inverse. More precisely, one rather has $\bo^{-1} \bo \ph = \ph + h$, where $h$ is a homogeneous solution $\bo h = 0$ which depends both on the choice of $\ph$ and $\bo^{-1}$, and thus $g_{\mu\nu}$\footnote{Consider for simplicity the operator $\pa_t^2$ and the following inverse 
\beq
(\pa^{-2} f)(t) \equiv \int_{-\infty}^{\infty} \ed t' \, G(t,t') f(t') \, , \hspace{1cm} G(t,t') = \te(t-t')\te(t'-t_0)(t-t') \, ,
\eeq
which gives $(\pa^{-2} \pa^2 f)(t) = f(t) - f(t_0) - f'(t_0)(t-t_0)$. Thus, with this definition, $\pa^{-2}$ is a left inverse $\pa^{-2} \pa^2 = {\rm id}$ only on the space of functions obeying $f(t_0) = f'(t_0) = 0$.}. This issue is relevant when computing the variation of $\bo^{-1}$ with respect to $g_{\mu\nu}$. Indeed, one applies $\de$ on $\bo \bo^{-1} = {\rm id}$ to get $\bo \de \bo^{-1} = - (\de \bo) \bo^{-1}$ and then one needs to cancel the $\bo$ {\it from the left} in order to isolate $\de \bo^{-1}$. For a given function $\ph$, there exists always a $\bo^{-1}$ such that $\bo^{-1} \bo \ph = \ph$\footnote{First pick an arbitrary $\bo^{-1}$ with Green's function $G$, in which case one has $\bo^{-1} \bo \ph = \ph + h[\ph]$ for some homogeneous solution $h[\ph]$. The desired inversion is then given by $(\bo^{-1})' \ph \equiv \bo^{-1} \ph - h[\ph]$ and, since $h[\ph]$ is a linear functional of $\ph$, this can be expressed as the convolution with a Green's function $G'$.}, so here $\de \bo^{-1} = - (\bo^{-1})' ( \de \bo ) \bo^{-1}$. However, $(\bo^{-1})'$ is not necessarily $\bo^{-1}$ and, most importantly, it depends on the function on which these operators act.

Fortunately, this issue is not relevant in this context because, as we have shown in the first paragraph, causality forces us to modify the result of the variation by hand: at the end of the computation we must anyway switch all the inverse d'Alembertians to (\ref{eq:retG})(\ref{eq:Gfchoice}). We can therefore perform the variation in a formal way, i.e. using only the symbol ``$\bo^{-1}$'' without caring about which inverses are meant in every instance, i.e. whether it is $(\bo^{-1})^T$ or any other $(\bo^{-1})'$ that appear\footnote{Alternatively, this amounts to working in the quotient space of scalar functions modulo homogeneous solutions of $\bo$, in which case $\bo^{-1}$ is uniquely defined and is both a right and a left inverse.}.

The important property of this ad hoc manipulation is that it preserves transversality $\na^{\mu} \De G_{\mu\nu} = 0$, which was the main advantage of starting with an action instead of modifying GR directly at the level of the equations of motion. As a consequence, the energy-momentum tensor is conserved, thus providing a sensible modification to classical GR. This additional prescription, that one must append to the variational principle, makes the relation between the final equations of motion and the action rather formal. It reflects in a concrete way what we stressed in the introduction, that the action should only be considered as merely a compact way to display the information of the equations of motion. Now that these issues have been clarified, we can perform the variation. Using 
\bea
\de R & = & \( R_{\mu\nu} + g_{\mu\nu} \bo - \na_{\mu} \na_{\nu} \) \de g^{\mu\nu} \, , \\
\(\de \bo\) \ph & = & \( \na_{\mu} \na_{\nu} \ph + \na_{\mu} \ph \na_{\nu} - \frac{1}{2}\, g_{\mu\nu} \na_{\ro} \ph \na^{\ro} \) \de g^{\mu\nu} \, ,
\eea
where $\ph$ is a scalar, and integrating by parts $\na$ and $\bo^{-1}$ a few times we arrive at (\ref{eq:EOM}) with
\bea
\De G_{\mu\nu} & = & \ti{m}^2 \[ - G_{\mu\nu} \frac{1}{\bo^2} R + \(\na_{\mu} \na_{\nu} - g_{\mu\nu} \bo \) \frac{1}{\bo^2} R + \frac{1}{4}\, g_{\mu\nu} \( \frac{1}{\bo}\, R \)^2 \] \nn \\
 & & - \frac{\ti{m}^2}{2} \( g_{\mu\ro} g_{\nu\si} + g_{\mu\si} g_{\nu\ro} - g_{\mu\nu} g_{\ro\si} \)  \( \na^{\ro} \frac{1}{\bo} R \) \( \na^{\si} \frac{1}{\bo^2} R \) \, , \label{eq:corr}
\eea
and $\ti{m} \equiv \al m$. As mentioned above, it is understood that here $\bo^{-1}$ is the retarded one obeying (\ref{eq:Gfchoice}). It is a simple task to check that $\na^{\mu} \De G_{\mu\nu} = 0$.

\subsection{Degree of freedom count} \label{sec:DOF}

In this section and the following one we use the same procedure which was used in \cite{DW2} for the Deser-Woodard model. We first go to the synchronous gauge
\beq \label{eq:syncgauge}
\ed s^2 = - \ed t^2 + h_{ij} \ed x^i \ed x^j \, ,
\eeq
and choose this time coordinate to be the time in the definition of the Green's function (\ref{eq:Gfchoice}). This is the synchronous time coordinate, i.e. the proper time of free-falling observers, which will become cosmic time when we go to consider cosmology. In GR, the dynamical equations of motion for the $h_{ij}$ fields are the ones of $g^{ij}$ since
\beq \label{eq:Gij}
G_{ij} = \frac{1}{2}\, \ddot{h}_{ij} - \frac{1}{2}\, h_{ij} \pa_t^2 \log h + \Ord ( \pa_t ) \, , \hspace{1cm} h \equiv \det h_{ij} \, ,
\eeq
while $G_{0\mu}$ starts at $\Ord ( \pa_t )$ and thus represents constraints on the initial data $h_{ij}(t_0), \dot{h}_{ij}(t_0)$. These constraints reduce the number of independent initial conditions of $h_{ij}$ and therefore determine the degrees of freedom of the theory\footnote{More rigorously, in the canonical (or ADM) formulation, one has $d+1$ first-class constraints which in turn generate $d+1$ gauge-transformations on phase-space, so that one can eliminate $2d + 2$ of the $d(d+1)$ canonical variables $h_{ij}, \pi^{ij}$. Therefore, the number of constraints is only half the number of eliminated variables, but since it comes with an equal number of gauge-fixing conditions (phase-space is even-dimensional), we can safely subtract twice the number of constraints to deduce the degrees of freedom.}. Consequently, the MM model will have the same degrees of freedom as GR if $\De G_{0\mu}(t_0) = 0$, which we show now. The terms $\sim \pa_t^n \bo^{-m}R$ with $n = 0, 1$ and $m = 1, 2$ are zero when $t = t_0$ because of the choice of Green's function (\ref{eq:Gfchoice}). Therefore, the only potentially dangerous term in $\De G_{0\mu}(t_0)$ is the one involving the operator $\( \na_0 \na_{\mu} - g_{0\mu} \bo\)$ which obviously does not contain second-order time-derivatives when evaluated on (\ref{eq:syncgauge}). As a consequence $\De G_{0\mu}(t_0) = 0$, so (\ref{eq:MM}) has the same $d^2 - d - 2$ independent initial conditions as GR and is thus also a theory of a massless spin-2 field in $d=3$.

\subsection{Stability of tensor modes} \label{sec:GWstab}

\begin{figure}
\begin{center}
\includegraphics[width=7cm]{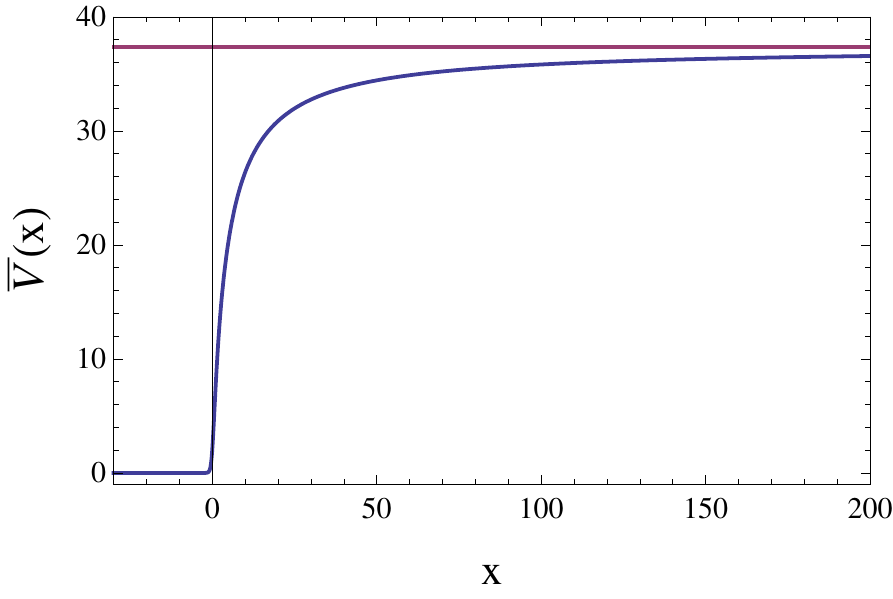} 
\caption{The background function $\bar{V}(x)$ (blue line) asymptotically reaching the value $1/3\ga$ (purple line) for the choice $\ga \approx 0.0089$ which reproduces the observed amount of dark energy.}
\label{fig:V} 
\end{center}
\end{figure}

As for any modification of GR (let alone GR itself), a rigorous proof of the stability of a given solution at the fully non-linear level is no walk in the park and goes of course well beyond the scope of this paper. It is however a lot simpler to study whether some necessary conditions for stability are fulfilled. As proposed in \cite{DW2}, an obvious such condition is that the correction $\De G_{ij}$ to the dynamical equations of motion
\beq
G_{ij} + \De G_{ij} = 8\pi G T_{ij} \, ,
\eeq
does not alter the behaviour of the propagating modes qualitatively. Simply put, if the sign of the coefficient in front of the kinetic term in (\ref{eq:Gij}) changes with respect to the one in GR, then one should expect that the graviton becomes a ghost\footnote{Note that only the first second-derivative term in (\ref{eq:Gij}) is a kinetic term for the propagating modes because $h$ is constrained by $G_{00}$.}. Writing the Ricci scalar as
\beq
R = -\bo \log h + \Ord(\pa_t) \, ,
\eeq
where here $\bo$ is the d'Alembertian in the scalar representation, shows that $\bo^{-1} R$ contains only the fields $h_{ij}$ and their first time-derivatives. The latter however are inside an integral of the form $\int^t$, since $\bo^{-1}$ is retarded, so one needs two time derivatives on $\bo^{-1} R$ to obtain second-order time derivatives of $h_{ij}$. Therefore, the only terms in (\ref{eq:corr}) containing second-order time derivatives are either those proportional to $G_{\mu\nu}$ or those with at least two derivatives acting on $\bo^{-1} R$. Obviously, only the first two terms are concerned. However, the second term can be written
\beq
\(\na_i \na_j - h_{ij} \bo \) \frac{1}{\bo^2} R = \na_i \na_j \frac{1}{\bo^2} R - h_{ij} \frac{1}{\bo} R \, ,
\eeq
so that it actually contains no second time-derivative. Therefore, the kinetic part of the dynamical equation is 
\beq \label{eq:kinetic}
\frac{1}{2} \( 1 - \frac{\ti{m}^2}{\bo^2} R \) \( \ddot{h}_{ij} - h_{ij} \pa_t^2 \log h \) + \Ord ( \pa_t ) = 8\pi G T_{ij} \, .
\eeq
We see that if $(1 - \ti{m}^2 \bo^{-2} R)$ becomes negative, then the graviton becomes a ghost. Most probably, one could always find a set of initial conditions such that $(1 - \ti{m}^2 \bo^{-2} R) < 0$ in some space-time region, so we will rather discuss the realistic cases. The fact that the mass term is chosen of the order of the Hubble scale today, along with the fact that the theory exhibits no vDVZ discontinuity \cite{MM, KM}, implies that $\ti{m}^2 \bo^{-2} R \ll 1$ at solar system length and time scales\footnote{In particular, we recover the stability under tensor excitations of the Schwarzschild space-time for small Schwarzschild radii compared to $m^{-1} \sim H_0^{-1}$, as it was already argued in \cite{KM}.}. The potential danger thus lies in cosmology. We must therefore study the background evolution of this quantity on the cosmological solution of interest, that is, $d=3$ with radiation and matter. To this end, translating to the definitions of \cite{DFKKM}, we find that
\beq \label{eq:ga}
1 - \frac{\ti{m}^2}{\bo^2} R = 1 - 3 \ga \bar{V} \, , \hspace{1cm} \ga \equiv \frac{m^2}{9 H_0^2}
\eeq
where the dimensionless quantity $\bar{V}$ is plotted in figure 1 of \cite{DFKKM}. We get that it starts growing today but that figure does not go far enough into the future to see whether the threshold value of $1/3\ga$ is crossed. We have therefore extended this window, only to find that this curve approaches $1/3\ga$ asymptotically (see figure \ref{fig:V}), so that $(1 - \ti{m}^2 \bo^{-2} R) > 0$ at all times. This proves the stability of gravitational waves at linear order in cosmological perturbation theory. 

It is important to note that the above stability arguments concerned the degrees of freedom of the theory only, i.e. the gravitational waves, since it focuses on the sign of the kinetic operator of these modes. However, it is possible that other excitations destabilize the solutions of interest and we see no easy way of dealing with these cases in this framework. For this reason, we now propose to discuss a local formulation of (\ref{eq:MM}), which displays all the potentially dangerous fields. It then allows one to address the questions of degree of freedom count and stability for all modes and in a more transparent way. Finally, apart from these theoretical advantages, the local formulation is also extremely useful in solving numerically the equations of motion as it transforms the system from an integro-differential to a differential one.

\section{Local formulation} \label{sec:loc}

Let us start by taking (\ref{eq:MM}) and integrating-in a scalar $\Phi$ so as to make the action local. A possible choice is
\beq \label{eq:loc1}
S_{\rm loc} =  \int \ed^{d+1} x\, \sqrt{-g} \[ M \Phi R + \frac{1}{2\ti{m}^2} \(\bo \Phi \)^2  \] \, .
\eeq
The equation of motion of $\Phi$ being
\beq 
\bo^2 \Phi = - M \ti{m}^2 R  \, ,
\eeq
the general solution is
\beq 
\Phi = \Phi_0 - M \frac{\ti{m}^2}{\bo^2} \, R  \, ,
\eeq
where $\Phi_0$ is a homogeneous solution $\bo^2 \Phi_0 = 0$  and $\bo^{-1}$ obeys (\ref{eq:retG})(\ref{eq:Gfchoice}). The freedom in choosing the initial conditions of $\Phi$ is reflected in the freedom of choosing $\Phi_0$. However, if we want (\ref{eq:loc1}) to reproduce $S_{\rm MM}$ when integrating-out $\Phi$, the only option is to choose the solution $\Phi_0 = M$, i.e.
\beq \label{eq:Phi_of_g} 
\Phi = M \(1 - \frac{\ti{m}^2}{\bo^2} \, R \)  \, .
\eeq
Alternatively, only this choice allows us to retrieve GR in the $m \to 0$ limit. This can equivalently be expressed as a choice of initial conditions for $\Phi$, which given our choice (\ref{eq:Gfchoice}) for $\bo^{-1}$, are given by\footnote{It is also easy to show that, if we integrate out $\Phi$ at the level of the equations of motion using (\ref{eq:Phi_of_g}), we retrieve the non-local equations (\ref{eq:EOM})(\ref{eq:corr}) that were obtained by varying the MM action and applying the ad hoc prescription for the $\bo^{-1}$.}
\beq \label{eq:inPhi}
\Phi(t_0) = M \, , \hspace{1cm} \dot{\Phi}(t_0) = \ddot{\Phi}(t_0) = \dddot{\Phi}(t_0) = 0 \, .
\eeq
Therefore, in non-local theories, integrating-in an auxiliary field in order to localize the action must be followed by additional data, the initial conditions of that field, if one wants to retrieve the original theory. This is quite peculiar because different choices for the initial conditions of $\Phi$ will lead to different theories, while for usual fields one rather gets different solutions of the same theory. Thus, (\ref{eq:inPhi}) are theory-level data, i.e. at the same level of information as the action (\ref{eq:loc1}) itself, and correspond to the prescription data for the Green's function (\ref{eq:Gfchoice}). Since there is no freedom in choosing the initial conditions of $\Phi$, the latter contributes zero to the degree of freedom count \cite{FMM1, DW2}. Consequently, (\ref{eq:loc1}) is {\it not} a scalar-tensor theory, as already noted in \cite{MM, FMM1, KM, DW2, B1, Kosh}.

This situation is in sharp contrast with higher-derivative theories where one also usually integrates-in auxiliary fields to lower the derivative order. In that case, however, the initial conditions of the auxiliary fields correspond to the initial conditions of the higher derivatives of the original fields, which now need less initial data\footnote{The simplest example is the passage to the canonical form of the action for a second-order theory where on integrates-in the conjugate momenta.}. Therefore, in that case, one simply distributes the initial data among more fields. The new fields have free initial conditions and therefore correspond to genuine degrees of freedom of the theory. 

Another important comparison one could make is with local actions exhibiting gauge symmetry. Indeed, there one also has fields that look dynamical, i.e. obey equations with time derivatives, but which do not correspond to degrees of freedom of the theory, i.e. whose initial conditions are constrained. For instance, in classical Yang-Mills theory these are the longitudinal modes, which are constrained by the Gauss constraint (and the gauge transformation it generates), while a GR example is the spacial volume density $h$, which is constrained by the Hamiltonian constraint (and the gauge transformation it generates). The crucial difference between these theories and our case is that the information for constraining these fields is inside the action itself, a direct consequence of gauge symmetry. It suffices to compute the equations of motion to see the constraints. Here, the concerned field is a scalar, so that there is obviously no gauge symmetry involved, and the constraints on the initial conditions have to be imposed by hand. 

Finally, note the obvious advantage of the localized formulation, which is that we do not need to invoke an ad hoc prescription for imposing causality anymore.

\subsection{Degree of freedom count}

Now that these important subtleties are understood, we see that we have obtained an alternative way for counting the degrees of freedom of the theory. We must first diagonalize the gravitational and scalar sectors by going to the Einstein frame. To that end, we note that $\Phi/M$ is actually the factor whose sign determines the stability of the gravitational field (see (\ref{eq:Phi_of_g}) and (\ref{eq:kinetic})). In fact, in the local formulation this conclusion is actually trivial since $\Phi$ is in front of the Ricci scalar in (\ref{eq:loc1}). Since in realistic scenarios $\Phi > 0$, we can consider the alternative variable $\Phi := M e^{\frac{\al \vph}{M}}$
\beq
S_{\rm loc} \to \int \ed^{d+1} x\, \sqrt{-g} \[ M^2 e^{\frac{\al \vph}{M}} R + \frac{1}{2 m^2} \, e^{\frac{2\al \vph}{M}} \( \bo \vph + \frac{\al}{M} \, \na_{\mu}\vph \na^{\mu} \vph \)^2  \] \, ,
\eeq
and then go to the Einstein frame
%\beq \label{eq:JtoE}
%g_{\mu\nu} = e^{2K}  \ti{g}_{\mu\nu} \, , \hspace{1cm} \sqrt{-g}\, R = e^{(d-1)K}\sqrt{-\ti{g}} \( \ti{R} - 2d \, \ti{\bo} K - d(d-1) \ti{\na}_{\mu} K \ti{\na}^{\mu} K \) \, ,
%\eeq
%with
%\beq
%K = -\frac{1}{d-1} \frac{\al\vph}{M} \, ,
%\eeq
\beq \label{eq:JtoE}
g_{\mu\nu} = e^{-\frac{2}{d-1} \frac{\al\vph}{M}}  \ti{g}_{\mu\nu} \, , 
\eeq
to get
\beq \label{eq:loc2}
S_{\rm loc} \to S'_{\rm loc} \equiv \int \ed^{d+1} x\, \sqrt{-\ti{g}} \[ M^2 \ti{R} - \frac{1}{2} \, \ti{\na}_{\mu} \vph \ti{\na}^{\mu} \vph + \frac{1}{2m^2} \, e^{\frac{2\vph}{\ti{M}}} \( \ti{\bo} \vph \)^2  \] \, ,
\eeq
where $\ti{M} \equiv 2 \frac{d-1}{d+1} \frac{M}{\al}$ and we have dropped a total derivative term\footnote{Note that the canonical normalization for $\vph$ unfortunately makes sense only for $m \neq 0$. If we want to keep track of the $m \to 0$ limit to GR, then we should rather use something like $\Phi := M e^{\frac{m \vph}{M^2}}$, so that for $m \to 0$ we get $\Phi \to M$. Here, the GR limit should therefore be understood as the combination $m \to 0, \vph \to 0$. \label{ft:GRlim}}. The initial conditions in terms of $\vph$ are now simply
\beq
\vph(t_0) = \dot{\vph}(t_0) = \ddot{\vph}(t_0) = \dddot{\vph}(t_0) = 0 \, .
\eeq
These being fixed, we get that the MM theory has the same degrees of freedom as GR since the only other field around is $\ti{g}_{\mu\nu}$ with the standard Einstein-Hilbert action. Finally, we can integrate-in a second auxiliary field $\psi$ in (\ref{eq:loc2}) so as to make the system second-order in derivatives 
\beq
S'_{\rm loc} \to \int \ed^{d+1} x\, \sqrt{-\ti{g}} \[ M^2 \ti{R} - \frac{1}{2}\, \ti{\na}_{\mu} \vph \ti{\na}^{\mu} \vph + \ti{\na}_{\mu} \vph \ti{\na}^{\mu} \psi  - \frac{m^2}{2} \, e^{-\frac{2\vph}{\ti{M}}} \psi^2  \] \, ,
\eeq
and diagonalize $\vph = \ph + \psi$ to get a canonical form
\beq \label{eq:loc3}
S'_{\rm loc} \to S''_{\rm loc} \equiv \int \ed^{d+1} x\, \sqrt{-\ti{g}} \[ M^2 \ti{R} - \frac{1}{2} \ti{\na}_{\mu} \ph \ti{\na}^{\mu} \ph + \frac{1}{2} \,\ti{\na}_{\mu} \psi \ti{\na}^{\mu} \psi  - \frac{1}{2}\,m^2 \psi^2 e^{-\frac{2(\ph + \psi)}{\ti{M}}}  \] \, ,
\eeq
with the initial conditions now being 
\beq
\ph(t_0) = \dot{\ph}(t_0) = \psi(t_0) = \dot{\psi}(t_0) = 0 \, .
\eeq
In this form, it is trivial to pinpoint the ghost mode.

\subsection{Stability}

In section \ref{sec:GWstab} we have seen that it is the condition $\Phi > 0$ which guarantees the stability of tensor modes, and here we see that it is also the same condition which allows us to diagonalize the action. We thus gain even more information than in the non-local analysis, namely, we have that {\it all} modes of $\ti{g}_{\mu\nu}$ have a healthy Lagrangian. The deviation from GR now lies in the extra scalar sector to which they couple, whose impact is not yet clear as far as stability is concerned. First of all, the special status of these fields, i.e. the fact that their initial conditions are frozen, implies that they must be handled in a special way in a stability analysis. Loosely speaking, a solution is stable if a ``small'' perturbation of its initial conditions yields a solution which is ``close enough'' to the original one\footnote{Note that this definition considers as a stable solution the case of a scalar field sitting in a local minimum of a potential even if the latter is not bounded from below. Also, the notion of ``close enough'' is of course subjective since it depends on the choice of a distance in field space and can be taken from either an absolute or a relative point of view.}. For a localizing auxiliary field however, perturbing the initial conditions of a solution amounts to changing the theory. So here we can only perturb the $d^2 - d -2$ independent initial conditions of $\ti{g}_{\mu\nu}$. Therefore, the only difference in the evolution of $\psi, \ph$ will come from the difference in the initial conditions of $\ti{g}_{\mu\nu}$. 

At first sight, this is a crucial remark because, quite generically, the auxiliary scalar in $\bo^{-1}$-based models contains a ghost, as we saw in our case. Such a field could destabilize a solution if its initial conditions were free to perturb. So the fact that these initial conditions are held fixed in using the stability criterion is what could allow the theory to remain classically stable. 

Unfortunately, there is a big caveat to this story. As already mentioned above, even if we perturb only the physical degrees of freedom, the evolution of the auxiliary fields will still change because of the interactions, i.e. non-linearities. So we can still explore different evolutions of the auxiliary fields, which in the case of the ghost might lead to a significant deviation from the original solution. The simplest example is the linearized theory (\ref{eq:loc3}) on the trivial background $\bar{g} = \et, \bar{\ph} = \bar{\psi} = 0$. Since the ghost has a healthy mass, its infrared modes $|k|< m$ diverge like $\sim e^{\sqrt{m^2-k^2} t}$ instead of oscillating. Without sources, fixing the initial conditions of the field to zero means that it will stay zero at all times, thus making the solution stable. Here we see that it can be crucial not to consider all the other spurious solutions of the larger scalar-tensor theory. However, if we now include the sources\footnote{Indeed, in the Einstein frame where the gravitational action is diagonal, the auxiliary scalars couple universally to matter.}, these will activate the field, whatever its initial conditions. Consider for instance the case where the source has compact support in time. The ghost field starts at zero and becomes non-zero as soon as the source is turned on. Then, when the latter is turned off, the evolution of the ghost is the same as in the free theory but having started with non-zero initial conditions\footnote{The same effect will take place when considering self-interactions as well. It also applies for tachyonic ghosts, whose solutions are plane waves, but non-linearities allow for an exchange of energy with an unbounded Hamiltonian and thus lead to an instability.}. At the end of the day, since the evolution of observables depends itself on the auxiliary scalars, we get a physical instability\footnote{See for example eqs. (3.5) to (3.7) of \cite{MM} in the $d+1$ harmonic decomposition, where the ghost field is $U$. To see this in the non-local formulation, see eq. (7) of \cite{M} where the model studied there has the same linearized theory as (\ref{eq:MM}). The trace of the equation of $h_{\mu\nu}$ in the harmonic gauge reads $\( \bo + m^2 \) h \sim T$.}.

Fortunately for the phenomenology of the model, starting with zero initial conditions, the divergence is visible only after a typical time lapse $\De t \sim m^{-1}$. Moreover, this concerns only wave-numbers of magnitude $k \sim m$, so Minkowski and Schwarzschild space-times are stable at space-time frequencies $\om, k \gg m \sim H_0$. This could have been expected from the absence of vDVZ discontinuity \cite{MM, KM}, that the physics are indistinguishable from GR at small scales. Moreover, since the amplitude of the ghost mode is small and well-behaved at such space-time scales, linear perturbation theory remains valid and we do not have to worry about the effect of non-linearities on the dynamics of the ghost\footnote{Indeed, in the $m \to 0$ limit we must retrieve GR, so the interaction of the ghost with gravity has to be controlled by some positive power of $m$, a fact which is not explicit in (\ref{eq:loc3}) where we have canonically normalized $\psi$ (see footnote \ref{ft:GRlim}).}.

We therefore turn ourselves to the FLRW solution with matter and radiation \cite{MM, DFKKM}. Considering the fact that: 1) $\psi$ has the wrong kinetic sign on all backgrounds, 2) we are in the presence of sources/non-linearities, and 3) now all the modes up to $k = 0$ must be considered, the stability of this solution is not guaranteed already at the level of linear perturbations. Nevertheless, given that these perturbations are known to be numerically well-behaved in the scalar sector \cite{DFKKM}, we must find out what mechanism stabilizes them. 

The answer lies in the time-dependence of the scale factor $a(t)$ in the DE era. From the numerical analysis of \cite{MM, DFKKM} we already know that $H$ grows so that the Hubble friction becomes more and more important. However, there is yet no clear reason of why this effect should dominate so as to tame even a ghost mode with non-trivial sources. Indeed, it may very well be the case that the time-dependence of other terms in the equation of motion of the ghost perturbation is strong enough to compensate that friction. As we will show in the next section, it is possible to compute analytically the asymptotic behaviour of $H$ in the future and we find that it actually reaches an infinite value at finite cosmic time $t$, the so-called ``big rip'' singularity. This then allows us to show that the Hubble friction will inevitably dominate at some point over any other effect and will thus drive the evolution of perturbations to a constant. We now proceed to compute the analytic solution of $a(t)$ in the DE era.

\section{Future singularity} \label{sec:bigrip}

\begin{figure}
\begin{center}
\includegraphics[width=7cm]{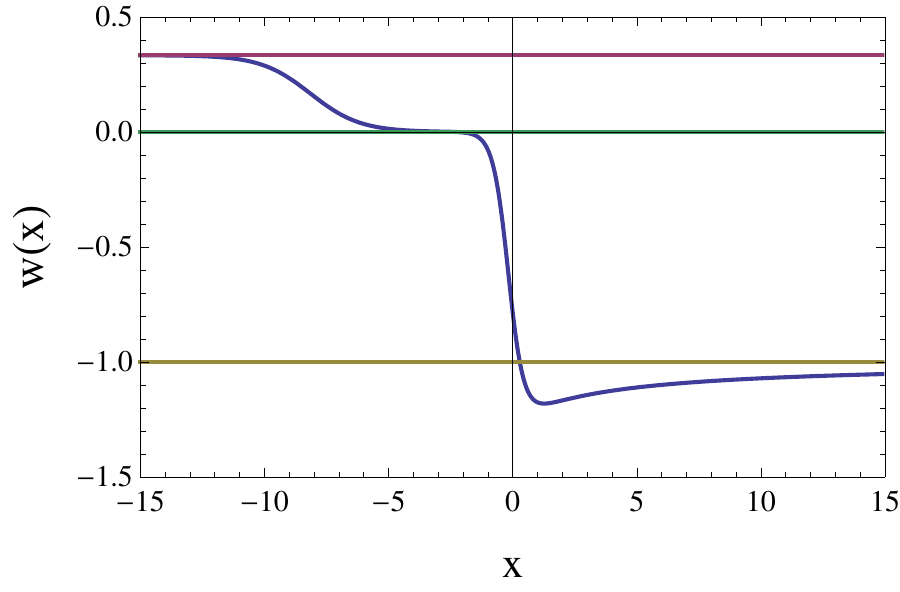} 
\caption{The total equation of state parameter $w$ as a function of $x$. The three plateaux correspond to the values for RD (purple), MD (green) and for the case of a cosmological constant (yellow).}
\label{fig:w} 
\end{center}
\end{figure}

\begin{figure}
\begin{center}
\includegraphics[width=14cm]{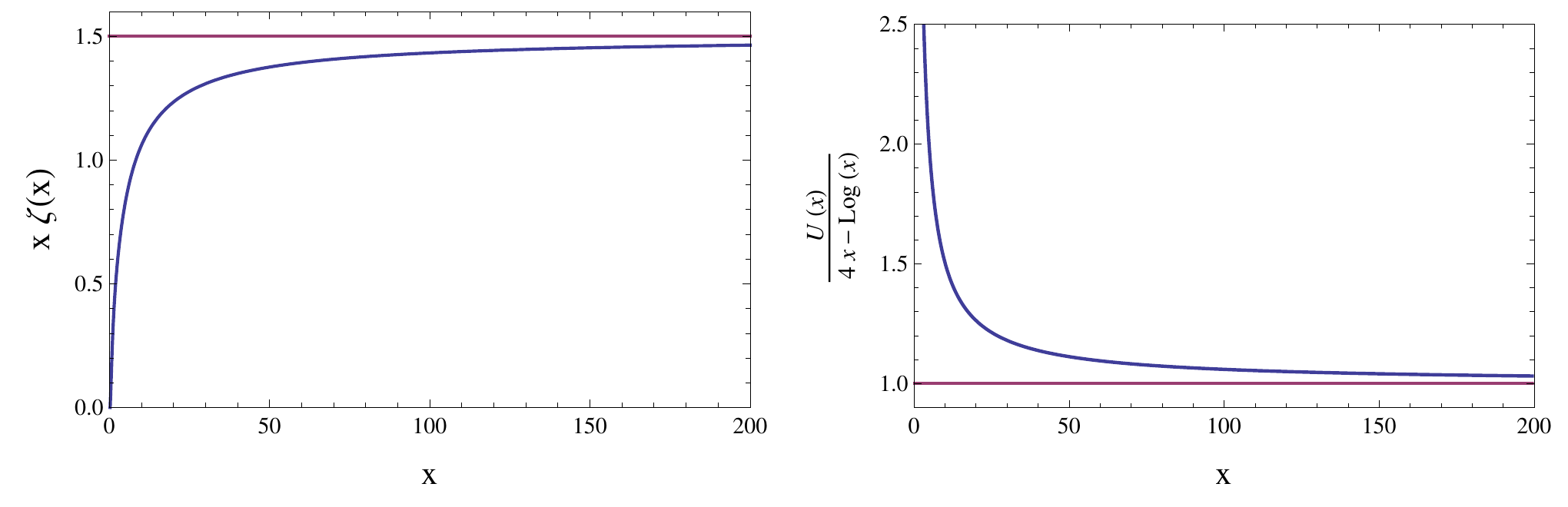} 
\caption{The functions $x \ze(x)$ (left) and $\frac{U(x)}{4x - \log x}$ (right) tending towards the constant values $3/2$ and $1$, respectively.}
\label{fig:zetaU} 
\end{center}
\end{figure}

Let us specialize to a flat FLRW background in $d=3$
\beq
\ed s^2 = - \ed t^2 + a^2(t) \ed \vec{x}^2 \, , \hspace{1cm} H \equiv \pa_t \log a \, ,
\eeq
and use the same definitions and localization procedure of \cite{DFKKM}, that is we define the following auxiliary scalars
\beq \label{eq:UVdef}
U \equiv -\bo^{-1} R \, , \hspace{1cm}  S \equiv - \bo^{-1} U = \bo^{-2} R \, ,
\eeq
whose initial conditions are all zero. We use the subscript ``$0$'' for quantities that are evaluated today and we set $a_0 = 1$. We then go to dimensionless variables, i.e. we define $V \equiv H_0^2 S$, $h \equiv H/H_0$ and use $x \equiv \log a$ as the time coordinate, with a prime to denote $\pa_x$. Note that this time extends indefinitely into the future for a monotonically increasing $a(t)$ even if $t$ is only defined up to a finite value. Moreover, consistency with our definitions requires that $h_0 = 1$ and this is achieved by appropriately tuning the only free parameter $\ga$ given in (\ref{eq:ga}). The first Friedmann equation (with $h^2$ isolated on the left-hand side), along with the equations for $U$ and $V$ can be found in equations (3.6) to (3.9) of \cite{DFKKM}, which we repeat here for convenience\footnote{Our numerical integration data are $\Om_R = 9.21 \times 10^{-5}$, $\Om_M = 0.3175$ \cite{P}, with the initial time being at $x_{\rm in} = -25$, matter-radiation equality at $x_{\rm eq} \approx -8.15$ and today at $x_0 = 0$. The value of the mass is fixed to $\ga = 0.0089247$ in order to get $h_0 = 1$ with a precision of five digits.\label{ft:numdata}}
\beq \label{eq:Fried}
h^2 = \frac{\Om + (\ga/4)  U^2}{1 + \ga \( -3 V' - 3 V + U' V' /2 \)} \, , 
\eeq
\bea
U'' + \(3 + \ze\) U' & = & 6\(2 + \ze\) \, , \label{eq:U} \\
V'' + \(3 + \ze\) V' & = & h^{-2} U \, ,
\eea
where
\bea 
\Om &\equiv & \Om_R e^{-4 x} + \Om_M e^{-3x} \, , \\
\ze & \equiv & \frac{h'}{h} = \frac{1}{2\( 1 - 3 \ga V \)} \[ h^{-2} \Om' + 3 \ga \( h^{-2} U + U' V' - 4 V' \) \] \, . \label{eq:zeta}
\eea
Note that in the last step we have simplified $\ze$ by using the other equations in order to get rid of high-order derivatives. Let us also define the total equation of state parameter $w \equiv p/ \ro$, where $\ro, p$ are the effective energy density and pressure, respectively\footnote{These are the quantities on the right-hand side of the Friedmann equations, when written in their usual form
\beq
H^2 = 8 \pi G \ro \, , \hspace{1cm} 2 \dot{H} + 3 H^2 = -8 \pi G p \, ,
\eeq
which thus capture any deviation from GR as an effective fluid source.}. As shown in \cite{FMM2}, $\ze$ and $w$ are related through
\beq \label{eq:wofzeta}
w = - \frac{2}{3}\, \ze - 1 \, .
\eeq
In figure \ref{fig:w} we see the usual first two plateau values that are $1/3$ and $0$ for RD and MD, respectively, and then we observe that $w$ tends towards $-1$ from below. 
Therefore, this is a so-called ``phantom dark energy''. The phenomenology of this type of dark energy was first considered in \cite{C,CKW}\footnote{For a study of the type of future singularities caused by phantom dark energy see \cite{NOT} and for the case where this occurs with the Deser-Woodard type of non-locality see \cite{BNOS}.} where it was realized that $w < -1$ would generically imply a future singularity at a finite time $t_{\rm rip}$
\beq \label{eq:bigrip}
\lim_{t \to t_{\rm rip}^-} a(t) = \infty \, .
\eeq
For constant $w$ this is easy to show. The continuity and first Friedmann equations read
\beq
\dot{\ro} + 3 H \( 1 + w \) \ro = 0 \, , \hspace{1cm} \dot{a} = a \sqrt{\frac{8\pi G}{3}\, \ro}  \, .
\eeq
The first gives $\ro = \ro_0 a^{-3(1+w)}$ and, plugging this in the second, we get
\beq
\dot{a} = H_0 \, a^{-\frac{3}{2} (1+w)+1} \, .
\eeq
The solution can be written as
\beq \label{eq:aofconstw}
a(t) = \[ - \frac{3}{2}\,H_0 \(1+w\) \( t_{\rm rip} - t \) \]^{\frac{2}{3(1+w)}} \, ,
\eeq  
where $t_{\rm rip}$ is the integration constant. Since $1+w < 0$, the bracket is positive, while the power is negative and we thus have (\ref{eq:bigrip}) indeed. The proof is less obvious for generic $w(t)$ as is the case here, but we can still find the explicit solution for $a(t)$ in the far future by proceeding perturbatively. Given (\ref{eq:wofzeta}) and figure \ref{fig:w}, we see that $\zeta$ tends towards zero as $x$ grows, so as a first approximation we can set $\zeta \approx 0$. This allows us to solve (\ref{eq:U}) analytically to find for $x \gg 1$
\beq \label{eq:U1}
U \approx 4 x  \, .
\eeq
However, if we plot $U/x$ we see that the function does not converge to a constant, so we need the next order correction. We know from section \ref{sec:GWstab} that $V$ tends towards a constant $1/3\ga$, and of course $\Om$ becomes negligible so, given (\ref{eq:U1}), one can approximate (\ref{eq:Fried}) and (\ref{eq:zeta}) by
\beq \label{eq:approx1}
h^2 \approx \frac{\ga U^2}{4\(1 - 3\ga V\)} \, , \hspace{1cm}  \ze \approx \frac{3 \ga U}{2 h^2 \( 1 - 3 \ga V \)}  \hspace{1cm} \Rightarrow \hspace{0.5cm} \zeta \approx \frac{6}{U}   \, .
\eeq
This gives us the first-order correction to $\zeta$
\beq 
\zeta \approx \frac{3}{2x} \, ,
\eeq
which is confirmed numerically in figure \ref{fig:zetaU}.
We then plug this improved $\zeta$ in the equation for $U$ and solve it again to get the next order correction for $x \gg 1$
\beq
U \approx 4x - \log x \, .
\eeq
We can now stop in solving $U$ perturbatively since we have gathered all terms that grow with $x$ as we can see in figure \ref{fig:zetaU}. However, this does not change the behaviour of $\zeta$ at $x \gg 1$ since, using (\ref{eq:approx1}), we now have
\beq
\zeta \approx \frac{6}{4x - \log x} = \frac{3}{2x} \sum_{k=0}^{\infty} \( \frac{\log x}{4 x} \)^k \approx \frac{3}{2x} \, .
\eeq
We can then solve $H' = \zeta H$ to find $H = (2/T) x^{3/2}$, for some positive constant $T$. To estimate the latter, we try to guess the asymptotic value of $x^{-3/2} h$ by going at large $x$ and find a good estimate in $x^{-3/2} h \to 0.09$, so we have that $T \approx 22 H_0^{-1}$. Finally, the equation for $a(t)$ is
\beq
\dot{a} = H a = \frac{2}{T} \( \log a \)^{3/2} a \, ,
\eeq
whose solution is
\beq \label{eq:DEera}
a(t) = \exp \[ \frac{T^2}{\( t_{\rm rip} - t \)^2} \] \, .
\eeq
At this point we must not forget that, since $H$ is growing in the DE (see figure 2 of \cite{MM}), which is an alternative way of defining a phantom dark energy, the curvature $R$ will eventually reach an energy scale where this phenomenological description ceases to be valid, so the region close to the singularity cannot be trusted.

\begin{figure}
\begin{center}
\includegraphics[width=14cm]{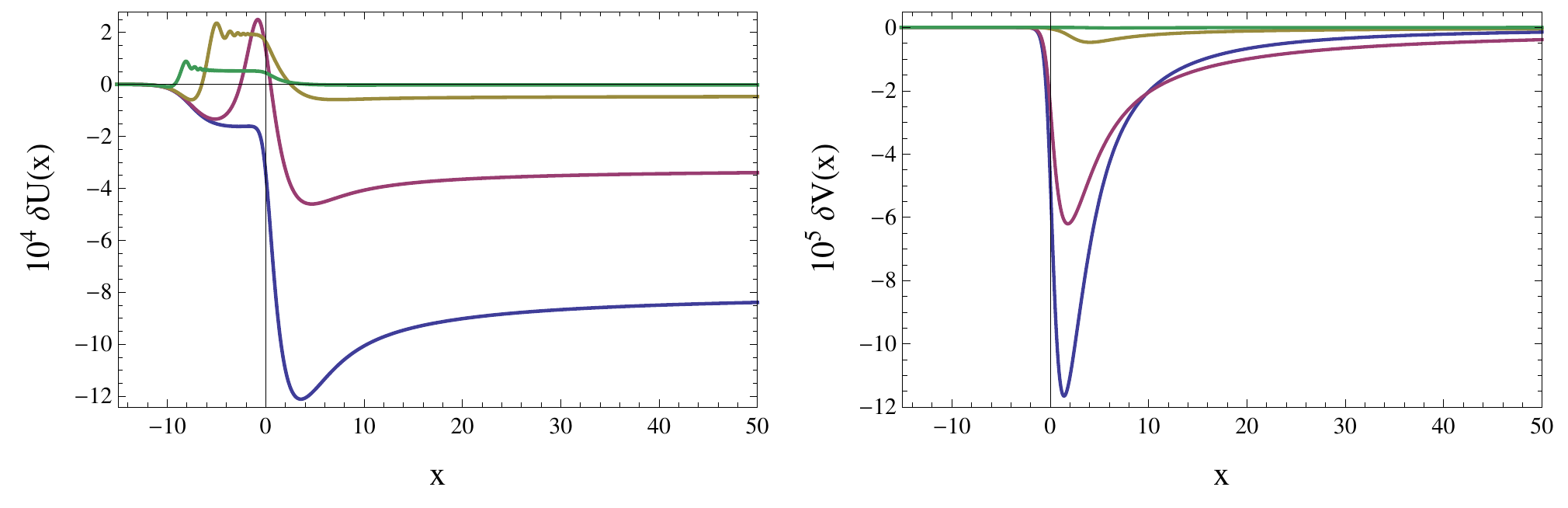} 
\caption{The linear perturbations of $U$ and $V$ as a function of $x$ for the modes $\ka = 5 \times 10^{-3}$ (blue), $\ka = 5 \times 10^{-2}$ (purple), $\ka = 5 \times 10^{-1}$ (brown), $\ka = 5$ (green).}
\label{fig:dUdV} 
\end{center}
\end{figure}

To conclude this section, we come back to the effect of Hubble friction on the perturbations. Given (\ref{eq:UVdef}), we have that the information of the perturbations of the ``canonical'' auxiliary scalars $\de \ph, \de \psi$ now lies in $\de U(t, \vec{k})$ and $\de V(t, \vec{k})$, so the ghost is in a combination of these fields\footnote{We refer the reader to \cite{DFKKM} for the corresponding equations of motion in section 4.1.}. In figure \ref{fig:dUdV} we have plotted these functions for several different values of comoving wave-number $\ka \equiv k / k_{\rm eq}$, where $k_{\rm eq} = a_{\rm eq} H_{\rm eq}$ is the comoving wave-number corresponding to the horizon scale at matter-radiation equality\footnote{For the numerical integration we have used the set-up of \cite{DFKKM}.}. Since $k_{\rm eq} \approx 42 H_0$, we have that the displayed choices of $\ka$ range from sub-horizon to super-horizon modes today and all of them tend to a constant for large $x \equiv \log a$. Incidentally, the same holds with respect to cosmic time $t$ and, in particular, they are smooth in the $t \to t_{\rm rip}^-$ limit.
%In \cite{DFKKM}, instead of focusing to $\de U, \de V$ themselves, the authors have chosen to treat the deviation from GR as an effective dark energy fluid and thus focused on the effective quantities $\ro_{\rm DE}, p_{\rm DE}, \te_{\rm DE}$ and $\si_{\rm DE}$ that are the energy density, pressure, velocity and anisotropic stress scalars, respectively. The conservation equation then leads to the evolution equation for the contrast $\de_{\rm DE} \equiv \de \ro_{\rm DE}/ \bar{\ro}_{\rm DE}$ which is eq. (6.9) of \cite{DFKKM}. In this description, the ``wrong'' relative sign appears in the fact that the sound speed squared $c_s^2$ is negative at all times, as shown in figure 16 of \cite{DFKKM}. The fact that $\de_{\rm DE}$ tends to zero in the future (figure 14 of \cite{DFKKM}) had already made the authors of \cite{DFKKM} deduce that the Hubble friction dominated the dynamics. 

Finally, one can be worried by the sudden jump $\de U, \de V$ as they enter the DE era, before being damped by the Hubble friction, especially in the case of large scales where the effect is the strongest. However, as shown in \cite{DFKKM}, this has no notable effect in the evolution of observable quantities such as the dark matter energy density or the Bardeen potentials.

\section{Conclusion} \label{sec:conclusion}

In this paper we have addressed the questions of stability of the Minkowski and FLRW solutions in the MM model (\ref{eq:MM}). We have shown that the tensor perturbations are well-behaved on both solutions. The potential danger lies in the ``hidden'' scalar sector of the theory which is unveiled in the localized formulation. These scalars look dynamical, in the sense that they obey equations that are second-order in time derivatives, but their initial conditions are fixed so that they do not correspond to degrees of freedom of the theory. Nevertheless, one of the two scalars is a ghost on all backgrounds and the non-linearities may drive an instability in the solutions of interest, whatever the initial conditions. Indeed, the Minkowski solution is found to be unstable under these excitations, but only for the infrared (cosmological) modes, i.e. out of the region of physical validity of this solution. For smaller scales, the absence of vDVZ discontinuity guarantees indeed that the physics become indistinguishable from GR. For cosmological scales it is the FLRW solution which is relevant and it has been shown (numerically) to be stable under {\it all} scalar excitations at the level of linear perturbation theory \cite{DFKKM}. This has been attributed to the domination of Hubble friction at late times, but no analytical proof was given. We have contributed to the understanding of this phenomenon by showing that the evolution of the scale factor leads to a big rip singularity. This implies that $H(t)$ will reach an infinite value at finite time, thus proving that Hubble friction will indeed inevitably dominate over any other effect and dilute any perturbation, including a ghost.

\section{Acknowledgements}

We are grateful to Stefano Foffa and Michele Maggiore for very useful comments and stimulating discussions. We would also like to thank the referee for pointing out the $\bo^{-1} \bo \neq {\rm id}$ issue. This work is supported by the Fonds National Suisse de la Recherche Scientifique.

\appendix

\section{Relation to quantum theory?}  \label{sec:Qstab}

The MM model is a classical theory defined by the equations (\ref{eq:retG})(\ref{eq:EOM})(\ref{eq:Gfchoice})(\ref{eq:corr}). The fact that it can be {\it formally} derived from a variational principle with some additional prescriptions opens the perspective for it being possibly related to a quantum field theory. A first natural question is therefore whether the MM model solutions can arise as the classical limit of some quantum theory. 

This issue has already been discussed in \cite{MM, M, DFKKM, FMM1} with arguments based on the linearized non-local formulation and the conclusion is that (\ref{eq:MM}) cannot be considered as a quantum action, since this would imply the presence of an ultra-light ghost and thus an unstable vacuum below the Hubble scale. Indeed, naively considering (\ref{eq:MM}) inside a path integral provides the following propagator for the linearized theory \cite{M, MM}
\bea 
\bra h_{\mu\nu}(k) h^*_{\ro\si}(k) \ket_{S_{\rm MM}} & = & -\frac{i}{k^2 - i \ep} \( \et_{\mu(\ro} \et_{\si)\nu} - \frac{1}{d-1}\, \et_{\mu\nu} \et_{\ro\si} \)  \\
 & &  - \frac{\et_{\mu\nu}\et_{\ro\si}}{d(d-1)} \( \frac{i}{k^2 - i \ep} + \frac{i}{-k^2 + m^2 - i \ep} \) + {\rm uncontracted}~ k{\rm 's}  \, , \nn \label{eq:prop}
\eea
where $h_{\mu\nu} \equiv M/\sqrt{2} \(g_{\mu\nu} - \et_{\mu\nu}\)$ is the canonically normalized field,
\beq
\bra \dots \ket_S \equiv \frac{\int D h \dots e^{i S[h]}}{\int D h \,e^{i S[h]}} \, ,
\eeq
and the $i \ep$ prescription chosen here is the one which makes the path integral converge, thus ensuring unitarity. We see that the propagator contains extra poles on top of the ones of the graviton, due to the presence of the $\bo^{-1}$ operator. Not surprisingly, these poles correspond precisely to the two auxiliary fields, i.e. a massless scalar and a ghost with a healthy mass $m$. The problem is that the path integral does not care about the interpretation of these extra poles as being not physical. They contribute just like the graviton ones to the residues in any quantum computation. Also, in the $m \to 0$ limit we do not retrieve GR since the propagators do not cancel because of the $i \ep$.

Here we wish to extend this discussion in the light of the local formulation we have introduced in this work. Indeed, since the MM model ultimately corresponds to a set of classical solutions $g_{\mu\nu}$, the formulation which is used to describe these solutions does not matter. The advantage of the local formulation is that it would allow in principle a cleaner road to a quantum theory. This is because considering a non-local action inside a path integral can hardly be motivated through (perturbative) canonical quantization, since one cannot define a Hamiltonian in the presence of time non-locality. 

In contrast, the local viewpoint provides a justification for using the linearized non-local action (\ref{eq:MM}) directly in a path integral, since one can then interpret it as the path integral of the linearized (\ref{eq:loc3}), where the scalars have been integrated-out\footnote{This equivalence does not hold at the non-linear level because the conformal transformation (\ref{eq:JtoE}) changes the measure $D g_{\mu\nu} D \Phi \sim e^{\be \int \ed^D x\, \vph} \, D \ti{g}_{\mu\nu} D \vph$, for some constant $\be$, assuming some sensible definition of the measures and integration domains.}. Indeed, using the linearized version of (\ref{eq:JtoE})
\beq
h_{\mu\nu} = \ti{h}_{\mu\nu} - \frac{\sqrt{2}\, \al}{d-1} \( \ph + \psi \) \et_{\mu\nu} \, ,
\eeq
we get that (\ref{eq:prop}) is actually
\beq \label{eq:lononlocrel}
\bra h_{\mu\nu}(k) h^*_{\ro\si}(k) \ket_{S_{\rm MM}} = \bra \ti{h}_{\mu\nu}(k) \ti{h}^*_{\ro\si}(k) \ket_{S''_{\rm loc}} + \frac{\et_{\mu\nu} \et_{\ro\si}}{d(d-1)} \bra \ph(k) \ph^*(k) + \psi(k) \psi^*(k) \ket_{S''_{\rm loc}} \, .
\eeq
However, in this case the fundamental theory is the scalar-tensor theory, with $\ph$ and $\psi$ considered as physical quantum fields and their poles correspond to physical particles. As already argued, the classical limit of this theory then contains many more solutions than the ones of the MM model since there are no constraints on the initial conditions of the scalars. Most importantly however, one of the scalar particles is a ghost, present on all backgrounds. At the classical level we saw that the perturbations of such a field are tamed by the Hubble friction, but at the quantum level this is a severe flaw because of vacuum decay \cite{CJM, Sb}. From an effective field theory point of view, ghosts are manageable only if their mass is higher than the cut-off, in which case the corresponding particles can never be produced. Here the mass $m$ is of the order of $H_0$ so this effective theory would only be valid for super-horizon physics and thus of little utility for our purposes. 

In this situation, an alternative option one could think of would be to promote the scalars to operators, but to somehow take into account their constrained nature so that we have no ghost particles nor spurious solutions in the classical limit. A priori, this may very well be a possibility since a similar construction is used in local gauge theories where the total Hilbert space includes the unphysical longitudinal polarization particles as well. However, thanks to gauge symmetry, the structure of the action and its consistent quantization (Faddeev-Popov fields, BRST symmetry) are such that the $S$-matrix is unitary in the physical Hilbert space, i.e. the subspace of transverse polarization particles. Starting with an ``in'' state which is made of only transverse polarized particles, we end up with an ``out'' state containing only such particles as well. The problem here is that the constraints have nothing to do with gauge symmetry, but are rather imposed by hand in order to describe the correct set of solutions. Therefore, starting with zero such particles the evolution will generically produce them and, projecting the ``out'' state onto the physical Hilbert space of pure gravitons, we will find that probability has not been conserved. Again, unitary evolution could only be achieved if the scalar particles are considered as physical, which as we saw is not an option. 

We therefore conclude that the simplest way to construct a unitary quantum field theory which would have the MM model solutions as its classical limit fails. Clearly, the difficulty lies in the fact that the prescriptions for the Green's functions or the scalars are hard to embed in a quantum context. Considering them as fixed theory-level data leaves no room for quantum uncertainty in the initial conditions of the scalars and these fields can therefore not be promoted to operators satisfying canonical commutation relations.

The next natural interpretation of the MM model one could think of, which would still relate it to quantum physics, is that it describes the dynamics of a vacuum expectation value. This is governed by the quantum effective action $\Ga$ in the Schwinger-Keldysh (or ``in-in'') formalism\footnote{This is the Legendre transform of the generating functional of some quantum field theory for the $\bra {\rm in} | \dots | {\rm in} \ket$ expectation values, and provides automatically causal equations of motion for $\bra {\rm in} | \hat{g}_{\mu\nu} | {\rm in} \ket$.}  \cite{S,BM1,BM2,K,J,V}. Indeed, non-local corrections are expected in $\Ga$ if the underlying theory $S$ contains massless particles \cite{DEM,TW} and these dominate over the local ones in the infrared. This fact has actually been one of the main theoretical justifications for studying non-local actions \cite{DW1, W, B2}. The problem with this interpretation however is that the non-local corrections one would expect from gravitational/matter loops look nothing like (\ref{eq:MM})\footnote{See \cite{DEM}, \cite{TW} and references therein for the corrections around Minkowski and de Sitter, respectively.}. Recently, in \cite{TW} the authors have actually shown that a large class of phenomenological models, including (\ref{eq:MM}), fails indeed to capture even qualitatively the non-local effects that are expected from quantum corrections on cosmological backgrounds. 

More generally, if the non-locality is a quantum correction in $\Ga$, then the appearance of a fixed mass scale $m$ would suggest a deviation from standard physics (GR + Standard Model), or at some more effective level, already in the original action $S$. With $m$ being ultra-light, we would expect to observe not only the corrections to gravity, i.e. the non-local term, but also other traces of this $m$-related new physics in $S$. We conclude that the mechanism which would generate the non-local correction (\ref{eq:corr}) to the dynamics of $g_{\mu\nu}$ or $\bra {\rm in} | \hat{g}_{\mu\nu} | {\rm in} \ket$ should therefore be more involved.

\end{document}

%% file: mydefs.tex
\newcommand{\beq}{\begin{equation}}
\newcommand{\eeq}{\end{equation}}
\newcommand{\bea}{\begin{eqnarray}}
\newcommand{\eea}{\end{eqnarray}}
\newcommand{\ben}{\begin{enumerate}}
\newcommand{\een}{\end{enumerate}}

\newcommand{\pa}{\partial}

\newcommand{\bo}{\square}
\newcommand{\na}{\nabla}
\newcommand{\ed}{{\rm d}}

\newcommand{\Ord}{{\cal O}}

\newcommand{\ti}{\tilde}

\renewcommand\({\left(}
\renewcommand\){\right)}
\renewcommand\[{\left[}
\renewcommand\]{\right]}
\newcommand{\bra}{\langle}
\newcommand{\ket}{\rangle}

\newcommand{\nn}{\nonumber}

\newcommand{\al}{\alpha}
\newcommand{\be}{\beta}
\newcommand{\ga}{\gamma}
\newcommand{\Ga}{\Gamma}
\newcommand{\de}{\delta}
\newcommand{\De}{\Delta}
\newcommand{\ep}{\epsilon}

\newcommand{\ze}{\zeta}
\newcommand{\et}{\eta}
\newcommand{\te}{\theta}

\newcommand{\ka}{\kappa}

\newcommand{\La}{\Lambda}
\newcommand{\ro}{\rho}
\newcommand{\si}{\sigma}

\newcommand{\ph}{\phi}
\newcommand{\vph}{\varphi}

\newcommand{\om}{\omega}
\newcommand{\Om}{\Omega}

%% file: MMstab2.bbl
\begin{thebibliography}{9}   

\bibitem{MM}
  M.~Maggiore and M.~Mancarella,
  %``Non-local gravity and dark energy,''
  Phys.\ Rev.\ D {\bf 90} (2014) 023005
  [arXiv:1402.0448 [hep-th]].
  %%CITATION = ARXIV:1402.0448;%%
  %12 citations counted in INSPIRE as of 20 Aug 2014

\bibitem{DFKKM}
  Y.~Dirian, S.~Foffa, N.~Khosravi, M.~Kunz and M.~Maggiore,
  %``Cosmological perturbations and structure formation in nonlocal infrared modifications of general relativity,''
  JCAP {\bf 1406} (2014) 033
  [arXiv:1403.6068 [astro-ph.CO]].
  %%CITATION = ARXIV:1403.6068;%%
  %3 citations counted in INSPIRE as of 01 Jul 2014

\bibitem{CKMT}
  A.~Conroy, T.~Koivisto, A.~Mazumdar and A.~Teimouri,
  %``Generalised Quadratic Curvature, Non-Local Infrared Modifications of Gravity and Newtonian Potentials,''
  arXiv:1406.4998 [hep-th].
  %%CITATION = ARXIV:1406.4998;%%
  %1 citations counted in INSPIRE as of 22 Jul 2014

\bibitem{BLHBP}
  A.~Barreira, B.~Li, W.~A.~Hellwing, C.~M.~Baugh and S.~Pascoli,
  %``Nonlinear structure formation in Nonlocal Gravity,''
  JCAP {\bf 1409} (2014) 031
  arXiv:1408.1084 [astro-ph.CO].
  %%CITATION = ARXIV:1408.1084;%%

\bibitem{DFKMP}
  Y.~Dirian, S.~Foffa, M.~Kunz, M.~Maggiore and V.~Pettorino,
  %``Non-local gravity and comparison with observational datasets,''
  arXiv:1411.7692 [astro-ph.CO].
  %%CITATION = ARXIV:1411.7692;%%

\bibitem{DW1}
  S.~Deser and R.~P.~Woodard,
  %``Nonlocal Cosmology,''
  Phys.\ Rev.\ Lett.\  {\bf 99} (2007) 111301
  [arXiv:0706.2151 [astro-ph]].
  %%CITATION = ARXIV:0706.2151;%%
  %115 citations counted in INSPIRE as of 11 Jun 2014

\bibitem{DW2}
  S.~Deser and R.~P.~Woodard,
  %``Observational Viability and Stability of Nonlocal Cosmology,''
  JCAP {\bf 1311} (2013) 036
  [arXiv:1307.6639 [astro-ph.CO]].
  %%CITATION = ARXIV:1307.6639;%%
  %22 citations counted in INSPIRE as of 13 Jun 2014

\bibitem{WD}
  C.~Deffayet and R.~P.~Woodard,
  %``Reconstructing the Distortion Function for Nonlocal Cosmology,''
  JCAP {\bf 0908} (2009) 023
  [arXiv:0904.0961 [gr-qc]].
  %%CITATION = ARXIV:0904.0961;%%
  %53 citations counted in INSPIRE as of 27 Aug 2014

\bibitem{K1}
  T.~Koivisto,
  %``Dynamics of Nonlocal Cosmology,''
  Phys.\ Rev.\ D {\bf 77} (2008) 123513
  [arXiv:0803.3399 [gr-qc]].
  %%CITATION = ARXIV:0803.3399;%%
  %80 citations counted in INSPIRE as of 14 Jul 2014

\bibitem{K2}
  T.~S.~Koivisto,
  %``Newtonian limit of nonlocal cosmology,''
  Phys.\ Rev.\ D {\bf 78} (2008) 123505
  [arXiv:0807.3778 [gr-qc]].
  %%CITATION = ARXIV:0807.3778;%%
  %51 citations counted in INSPIRE as of 14 Jul 2014

\bibitem{DP1}
  S.~Park and S.~Dodelson,
  %``Structure formation in a nonlocally modified gravity model,''
  Phys.\ Rev.\ D {\bf 87} (2013) 024003
  [arXiv:1209.0836 [astro-ph.CO]].
  %%CITATION = ARXIV:1209.0836;%%
  %26 citations counted in INSPIRE as of 27 Aug 2014

\bibitem{DP2}
  S.~Dodelson and S.~Park,
  %``Nonlocal Gravity and Structure in the Universe,''
  Phys.\ Rev.\ D {\bf 90} (2014) 043535
  [arXiv:1310.4329 [astro-ph.CO]].
  %%CITATION = ARXIV:1310.4329;%%
  %15 citations counted in INSPIRE as of 24 Sep 2014

\bibitem{W}
  R.~P.~Woodard,
  %``Nonlocal Models of Cosmic Acceleration,''
  Found.\ Phys.\  {\bf 44} (2014) 213
  [arXiv:1401.0254 [astro-ph.CO]].
  %%CITATION = ARXIV:1401.0254;%%
  %15 citations counted in INSPIRE as of 21 Aug 2014

\bibitem{JMM}
  M.~Jaccard, M.~Maggiore and E.~Mitsou,
  %``Nonlocal theory of massive gravity,''
  Phys.\ Rev.\ D {\bf 88} (2013) 4,  044033
  [arXiv:1305.3034 [hep-th]].
  %%CITATION = ARXIV:1305.3034;%%
  %18 citations counted in INSPIRE as of 11 Jun 2014

\bibitem{MT}
  L.~Modesto and S.~Tsujikawa,
  %``Non-local massive gravity,''
  Phys.\ Lett.\ B {\bf 727} (2013) 48
  [arXiv:1307.6968 [hep-th]].
  %%CITATION = ARXIV:1307.6968;%%
  %13 citations counted in INSPIRE as of 11 Jun 2014

\bibitem{M}
  M.~Maggiore,
  %``Phantom dark energy from non-local infrared modifications of General Relativity,''
  Phys.\ Rev.\ D {\bf 89} (2014) 043008
  [arXiv:1307.3898 [hep-th]].
  %%CITATION = ARXIV:1307.3898;%%
  %12 citations counted in INSPIRE as of 11 Jun 2014

\bibitem{FMM2}
  S.~Foffa, M.~Maggiore and E.~Mitsou,
  %``Cosmological dynamics and dark energy from nonlocal infrared modifications of gravity,''
  Int.\ J.\ Mod.\ Phys.\ A {\bf 29} (2014) 1450116
  [arXiv:1311.3435 [hep-th]].
  %%CITATION = ARXIV:1311.3435;%%
  %18 citations counted in INSPIRE as of 23 Sep 2014

\bibitem{KM}
  A.~Kehagias and M.~Maggiore,
  %``Spherically symmetric static solutions in a nonlocal infrared modification of General Relativity,''
  JHEP {\bf 1408} (2014) 029
  [arXiv:1401.8289 [hep-th]].
  %%CITATION = ARXIV:1401.8289;%%
  %8 citations counted in INSPIRE as of 20 Aug 2014

\bibitem{NT}
  S.~Nesseris and S.~Tsujikawa,
  %``Cosmological perturbations and observational constraints on nonlocal massive gravity,''
  Phys.\ Rev.\ D {\bf 90} (2014) 024070
  [arXiv:1402.4613 [astro-ph.CO]].
  %%CITATION = ARXIV:1402.4613;%%
  %7 citations counted in INSPIRE as of 20 Aug 2014

\bibitem{B1}
  A.~O.~Barvinsky,
  %``Serendipitous discoveries in nonlocal gravity theory,''
  Phys.\ Rev.\ D {\bf 85} (2012) 104018
  [arXiv:1112.4340 [hep-th]].
  %%CITATION = ARXIV:1112.4340;%%
  %21 citations counted in INSPIRE as of 11 Jul 2014

\bibitem{P}
  P.~A.~R.~Ade {\it et al.}  [Planck Collaboration],
  %``Planck 2013 results. XVI. Cosmological parameters,''
  arXiv:1303.5076 [astro-ph.CO].
  %%CITATION = ARXIV:1303.5076;%%
  %1929 citations counted in INSPIRE as of 11 Jun 2014

\bibitem{SNe}
  M.~Betoule {\it et al.}  [SDSS Collaboration],
  %``Improved cosmological constraints from a joint analysis of the SDSS-II and SNLS supernova samples,''
  %Submitted to: Astron.Astrophys.
  [arXiv:1401.4064 [astro-ph.CO]].
  %%CITATION = ARXIV:1401.4064;%%
  %10 citations counted in INSPIRE as of 11 Jun 2014

\bibitem{FMM1}
  S.~Foffa, M.~Maggiore and E.~Mitsou,
  %``Apparent ghosts and spurious degrees of freedom in non-local theories,''
  Phys.\ Lett.\ B {\bf 733} (2014) 76
  [arXiv:1311.3421 [hep-th]].
  %%CITATION = ARXIV:1311.3421;%%
  %16 citations counted in INSPIRE as of 22 Jul 2014

\bibitem{JNOSTZ}
  S.~Jhingan, S.~Nojiri, S.~D.~Odintsov, M.~Sami, I.~Thongkool and S.~Zerbini,
  %``Phantom and non-phantom dark energy: The Cosmological relevance of non-locally corrected gravity,''
  Phys.\ Lett.\ B {\bf 663} (2008) 424
  [arXiv:0803.2613 [hep-th]].
  %%CITATION = ARXIV:0803.2613;%%
  %63 citations counted in INSPIRE as of 27 Aug 2014

\bibitem{NO1}
  S.~'i.~Nojiri and S.~D.~Odintsov,
  %``Modified non-local-F(R) gravity as the key for the inflation and dark energy,''
  Phys.\ Lett.\ B {\bf 659} (2008) 821
  [arXiv:0708.0924 [hep-th]].
  %%CITATION = ARXIV:0708.0924;%%
  %112 citations counted in INSPIRE as of 14 Jul 2014

\bibitem{Kosh}
  N.~A.~Koshelev,
  %``Comments on scalar-tensor representation of nonlocally corrected gravity,''
  Grav.\ Cosmol.\  {\bf 15} (2009) 220
  [arXiv:0809.4927 [gr-qc]].
  %%CITATION = ARXIV:0809.4927;%%
  %23 citations counted in INSPIRE as of 14 Jul 2014

\bibitem{CENO}
  S.~Capozziello, E.~Elizalde, S.~'i.~Nojiri and S.~D.~Odintsov,
  %``Accelerating cosmologies from non-local higher-derivative gravity,''
  Phys.\ Lett.\ B {\bf 671} (2009) 193
  [arXiv:0809.1535 [hep-th]].
  %%CITATION = ARXIV:0809.1535;%%
  %68 citations counted in INSPIRE as of 14 Jul 2014

\bibitem{NOSZ}
  S.~Nojiri, S.~D.~Odintsov, M.~Sasaki and Y.~l.~Zhang,
  %``Screening of cosmological constant in non-local gravity,''
  Phys.\ Lett.\ B {\bf 696} (2011) 278
  [arXiv:1010.5375 [gr-qc]].
  %%CITATION = ARXIV:1010.5375;%%
  %26 citations counted in INSPIRE as of 27 Aug 2014

\bibitem{NO2}
  S.~Nojiri and S.~D.~Odintsov,
  %``Unified cosmic history in modified gravity: from F(R) theory to Lorentz non-invariant models,''
  Phys.\ Rept.\  {\bf 505} (2011) 59
  [arXiv:1011.0544 [gr-qc]].
  %%CITATION = ARXIV:1011.0544;%%
  %694 citations counted in INSPIRE as of 27 Aug 2014

\bibitem{BNOS}
  K.~Bamba, S.~Nojiri, S.~D.~Odintsov and M.~Sasaki,
  %``Screening of cosmological constant for De Sitter Universe in non-local gravity, phantom-divide crossing and finite-time future singularities,''
  Gen.\ Rel.\ Grav.\  {\bf 44} (2012) 1321
  [arXiv:1104.2692 [hep-th]].
  %%CITATION = ARXIV:1104.2692;%%
  %21 citations counted in INSPIRE as of 27 Aug 2014

\bibitem{C}
  R.~R.~Caldwell,
  %``A Phantom menace?,''
  Phys.\ Lett.\ B {\bf 545} (2002) 23
  [astro-ph/9908168].
  %%CITATION = ASTRO-PH/9908168;%%
  %1685 citations counted in INSPIRE as of 03 Jul 2014

\bibitem{CKW}
  R.~R.~Caldwell, M.~Kamionkowski and N.~N.~Weinberg,
  %``Phantom energy and cosmic doomsday,''
  Phys.\ Rev.\ Lett.\  {\bf 91} (2003) 071301
  [astro-ph/0302506].
  %%CITATION = ASTRO-PH/0302506;%%
  %1114 citations counted in INSPIRE as of 03 Jul 2014

\bibitem{TW}
  N.~C.~Tsamis and R.~P.~Woodard,
  %``A Caveat on Building Nonlocal Models of Cosmology,''
  JCAP {\bf 1409} (2014) 008
  [arXiv:1405.4470 [astro-ph.CO]].
  %%CITATION = ARXIV:1405.4470;%%
  %3 citations counted in INSPIRE as of 24 Sep 2014

\bibitem{NOT}
  S.~Nojiri, S.~D.~Odintsov and S.~Tsujikawa,
  %``Properties of singularities in (phantom) dark energy universe,''
  Phys.\ Rev.\ D {\bf 71} (2005) 063004
  [hep-th/0501025].
  %%CITATION = HEP-TH/0501025;%%
  %600 citations counted in INSPIRE as of 27 Aug 2014

\bibitem{CJM}
  J.~M.~Cline, S.~Jeon and G.~D.~Moore,
  %``The Phantom menaced: Constraints on low-energy effective ghosts,''
  Phys.\ Rev.\ D {\bf 70} (2004) 043543
  [hep-ph/0311312].
  %%CITATION = HEP-PH/0311312;%%
  %443 citations counted in INSPIRE as of 20 Jun 2014

\bibitem{Sb}
  F.~Sbis\`a,
  %``Classical and quantum ghosts,''
  arXiv:1406.4550 [hep-th].
  %%CITATION = ARXIV:1406.4550;%%

\bibitem{S}
  J.~S.~Schwinger,
  %``Brownian motion of a quantum oscillator,''
  J.\ Math.\ Phys.\  {\bf 2} (1961) 407.
  %%CITATION = JMAPA,2,407;%%
  %1022 citations counted in INSPIRE as of 16 Jun 2014

\bibitem{BM1}
  P.~M.~Bakshi and K.~T.~Mahanthappa,
  %``Expectation value formalism in quantum field theory. 1.,''
  J.\ Math.\ Phys.\  {\bf 4} (1963) 1.
  %%CITATION = JMAPA,4,1;%%
  %258 citations counted in INSPIRE as of 16 Jun 2014

\bibitem{BM2}
  P.~M.~Bakshi and K.~T.~Mahanthappa,
  %``Expectation value formalism in quantum field theory. 2.,''
  J.\ Math.\ Phys.\  {\bf 4} (1963) 12.
  %%CITATION = JMAPA,4,12;%%
  %234 citations counted in INSPIRE as of 16 Jun 2014

\bibitem{K}
  L.~V.~Keldysh,
  %``Diagram technique for nonequilibrium processes,''
  Zh.\ Eksp.\ Teor.\ Fiz.\  {\bf 47} (1964) 1515
   [Sov.\ Phys.\ JETP {\bf 20} (1965) 1018].
  %%CITATION = ZETFA,47,1515;%%
  %997 citations counted in INSPIRE as of 16 Jun 2014

\bibitem{J}
  R.~D.~Jordan,
  %``Effective Field Equations for Expectation Values,''
  Phys.\ Rev.\ D {\bf 33} (1986) 444.
  %%CITATION = PHRVA,D33,444;%%
  %287 citations counted in INSPIRE as of 29 Jul 2014

\bibitem{V}
  G.~A.~Vilkovisky,
  %``Expectation values and vacuum currents of quantum fields,''
  Lect.\ Notes Phys.\  {\bf 737} (2008) 729
  [arXiv:0712.3379 [hep-th]].
  %%CITATION = ARXIV:0712.3379;%%
  %8 citations counted in INSPIRE as of 29 Jul 2014

\bibitem{DEM}
  J.~F.~Donoghue and B.~K.~El-Menoufi,
  %``Non-local quantum effects in cosmology 1: Quantum memory, non-local FLRW equations and singularity avoidance,''
  Phys.\ Rev.\ D {\bf 89} (2014) 104062
  [arXiv:1402.3252 [gr-qc]].
  %%CITATION = ARXIV:1402.3252;%%
  %2 citations counted in INSPIRE as of 23 Jul 2014

\bibitem{B2}
  A.~O.~Barvinsky,
  %``Aspects of Nonlocality in Quantum Field Theory, Quantum Gravity and Cosmology,''
  arXiv:1408.6112 [hep-th].
  %%CITATION = ARXIV:1408.6112;%%


\end{thebibliography}
